# NEW INTERNAL STRESS DRIVEN ON-CHIP MICROMACHINES FOR EXTRACTING MECHANICAL PROPERTIES OF THIN FILMS

*D. Fabrègue[1,2], N. André[1,3], T. Pardoen[1,2], J.-P. Raskin[1,3]*

[1] Research Center in Micro and Nanoscopic Materials and Electronic Devices (CeRMiN)
Université catholique de Louvain, B-1348 Louvain-la-Neuve, Belgium
[2] Department of Materials Science and Process, IMAP, Université catholique de Louvain
B-1348 Louvain-la-Neuve, Belgium
[3] Department of Electrical Engineering, EMIC, Université catholique de Louvain
B-1348 Louvain-la-Neuve, Belgium

**ABSTRACT**

A new concept of micromachines has been developed for measuring the mechanical properties of thin metallic films. The actuator is a beam undergoing large internal stresses built up during the deposition process. Al thin films are deposited partly on the actuator beam and on the substrate. By etching the structure, the actuator contracts and pulls the Al film. Full stress strain curves can be generated by designing a set of micromachines with various actuator lengths. In the present study, the displacements have been measured by scanning electronic microscopy. The stress is derived from simple continuum mechanics relationships. The tensile properties of Al films of various thicknesses have been tested. A marked increase of the strength with decreasing film thickness is observed.

## 1. INTRODUCTION

Over the last few years, the interest for building "lab-on-chip" in order to test various physical properties of fluids, solids and biological species has become very popular. Among others, "labs-on-chip" concepts can be extremely useful for measuring the mechanical properties of small materials samples, especially thin films or thin beams made of single or multi-layers. Proper estimation of the mechanical response of thin metallic layers when thermal or mechanical stresses are applied is of a prime importance to assess the integrity of microsystems.

Several concepts of micromachines and loading set-ups have been proposed in the recent literature. A simple way to measure the mechanical properties of thin film is nano-indentation [1]. This technique allows determining the hardness of the film as well as the Young's modulus. But as the film gets thinner, the effect of the substrate becomes non negligible. Pile-up and sink-in effects resulting from the plastic behavior of the film during nano-indentation involve changes of the contact area and have to be taken into account in the calculation of the film properties [2]. The nano-indentor has also been used to bend free standing cantilevers of metallic glasses [3] silicon nitride [4] or silicon [5] in order to extract elastic properties of thin films. Nevertheless, the technique is very sensitive to film thickness and also suffers from difficulties in accounting for boundary conditions. Another limit comes from the inhomogeneous strain distribution and the difficulty to extract intrinsic material properties. Espinosa *et al.* [6] used a freestanding film and nano-indentor to impose displacement up to plastic yielding range. Bending of free standing beams can also be supplied by ultrasonic excitation [7] but only elastic properties can be extracted. Another bending technique is based on the measurement of the wafer curvature. The thin film is deposited at high temperature. The difference of thermal expansion coefficients between the film and the substrate induces a curvature of the wafer when the temperature decreases. By varying the temperature, the mechanical properties of the film can be measured [8]. This curvature technique has several limitations: firstly, it relies on the assumptions that the thin film accommodates entirely the lattice mismatch; secondly the bending cannot be higher than the half thickness of the substrate. At last, viscous effects related to temperature changes can make the analysis of the results more complex. The bulge test constitutes another bending test [9]. In this case a freestanding film is deformed under pressure imposed by a compressed gas or liquid. This test is limited to thin films with tensile residual stresses and boundary conditions have to be taken into account to determine the stress and the strain in the film.

Tensile testing is the most direct method to extract stress strain responses. Tensile testing stages have already been developed to measure the mechanical response of thin films. Piezoelectric actuator [10] or an electrostatic force gripping system [11] have been designed to pull the thin films. However in these cited cases, problems can occur due to sample preparation, handling and mounting. To avoid these problems, some authors have used microelectromechanical systems to test thin films. The material to be tested is directly deposited and released in the form of a freestanding tensile sample, linked to microelectromechanical system in order to provide the displacement. The microelectromechanical system could be a bulk piezoelectric actuator as in [12], an electrostatic actuator [13] or an electrothermal one [14]. These testing methods are promising and generate the mechanical response of micro- and nanoscale device. They require complex microelectromechanical systems in order to impose the loading to the sample.

The concept presented here constitutes a change of perspective in the mechanical testing of submicron scale films. The idea is to multiply elementary testing stages to obtain the stress and the strain rather than using a more complex multipurpose machine. In the specific implementation described in this paper, thermal stresses due to the difference in thermal expansion coefficients between a silicon nitride film (called the actuator) and the Si wafer substrate provide, after release, the actuation system. The tested film is connected to both the actuator beam and the substrate (Figures 1, 2 and 3). By chemically releasing the structure, the actuator beam contracts and imposes a displacement to the film. The measurement of the displacement of the actuator beam provides the value of the strain applied to the film. Subtracting this displacement from the displacement of a free actuator beam leads directly to the value of the stress. Different strains can be imposed to the film by using various combinations of film and beam lengths in order to generate full stress strain curves up to the fracture strain.

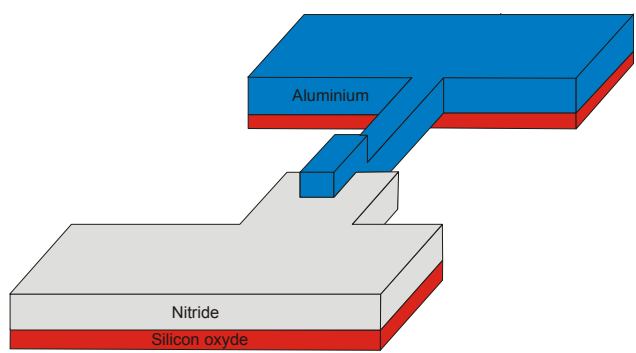

Figure 1: 3-D scheme of an elementary tensile testing stage.

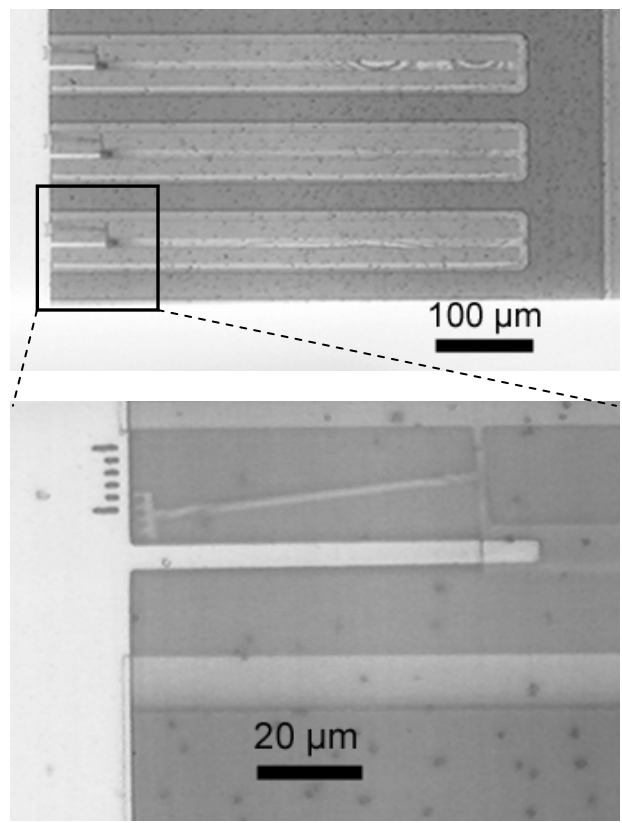

Figure 2: Design of the tensile test on a 500 nm aluminum film.

## 2. FABRICATION PROCESS

A micromachine for pulling micro or nano specimens in tension is shown in Fig. 2: a beam of silicon nitride (actuator) applies a displacement to an aluminum beam (tested film) when it contracts after release. Owing to conceptual simplicity of these micromachines, the fabrication process is relatively straightforward and therefore reliable. The full microfabrication process requires only three film depositions and two photolithographic steps to build the microstructures (Figure 3). First, a 1 μm-thick PECVD oxide (sacrificial layer) is deposited on top of a bulk silicon wafer (handling substrate). Secondly, a LPCVD silicon nitride film of 250 or 500 nm is deposited at around 800°C and patterned by dry etching ($CHF_3$ and $O_2$ plasma) to define the actuator beam. Then, an aluminum layer (250 nm-thick for the tests reported here) is evaporated in an E-gun vacuum system at a temperature of around 150°C and patterned in a chlorine-based plasma ($Cl_2$ and $CCl_4$). Finally, the release of both beams (actuator and tested material) is performed by wet etching of the PECVD silicon oxide (sacrificial layer). This wet etching is performed in a mixture of concentrated fluoridric acid (HF 73%) and isopropanol (IPA), HF:IPA=1:1. The rinsing is made in IPA. The use of water must be avoided because mixtures of HF and water etch aluminum. The micromachines are therefore placed in a CPD (Critical Point Dryer) machine to dry the samples without occurrence of stiction problems between the released beams and the substrate surface.

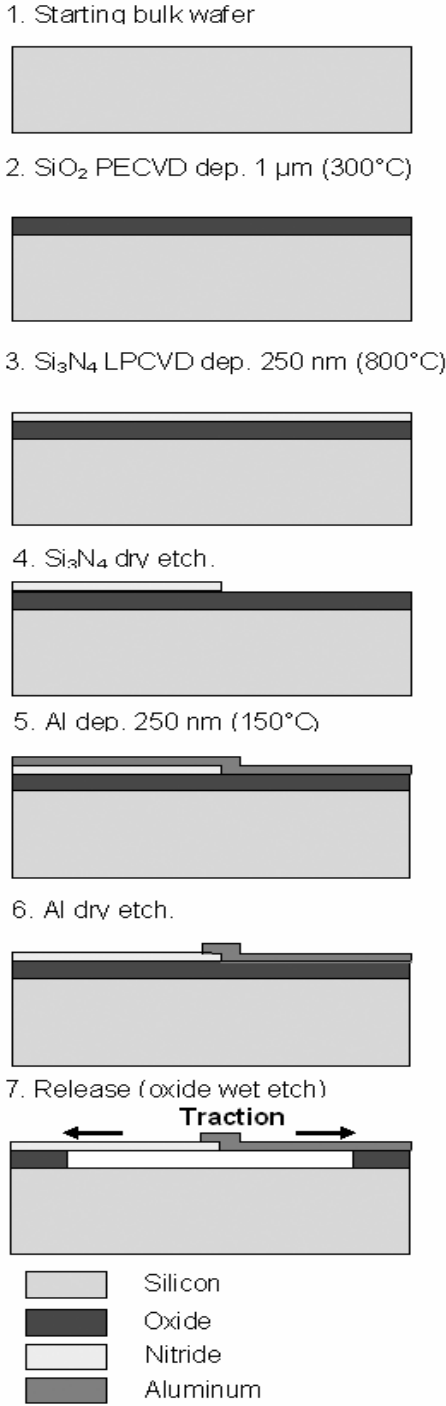

Figure 3: Scheme of the fabrication process.

# 3. RESULTS

## 3.1. Mechanical response of thin films

In order to determine precisely the deformation of the aluminum film and of the nitride actuator, nitride verniers have been designed to magnify the displacements. Due to out of plane bending of the vernier, these verniers did not always lead to accurate measurements of the displacement. Displacements have thus been also measured using Scanning Electron Microscopy. The observations have been carried out just after release in order to avoid creep and other relaxation effects. The displacement of the aluminum and the actuator beams have been measured on high magnification micrographs. The displacement of the aluminum beam is directly related to the logarithmic strain (which is a more physical measure than the engineering strain when the strains are large) through:

$$\varepsilon_{al} = \ln\left(\frac{l_{al}}{l_{al}^0}\right) = \ln\left(\frac{l_{al}^0 - \Delta l_{al}}{l_{al}^0 - \Delta l_{al}^{free}}\right) \quad (1)$$

with $l_{al}^0$ the initial length of the aluminum beam at room temperature, $l_{al}$ the length of the aluminum beam after release, $\Delta l_{al}$ the measured displacement and $\Delta l_{al}^{free}$ is the displacement for a beam without link to the actuator. The initial length of the aluminum beam $l_{al}^0$ is the length deposited corrected by the thermal strain involves by the temperature of the deposition:

$$l_{al}^0 = l_{al}^d (1 - \alpha_{Si} \Delta T) \quad (2)$$

with $l_{al}^d$ the length of the aluminum beam deposited, $\alpha_{Si}$ the thermal expansion coefficient of the silicon wafer and $\Delta T$ the difference of temperature between the deposition and the room temperatures. The deposition temperature has been measured to be equal to about 150°C. Moreover, calibration of the term $\alpha_{Si}\Delta T$ can be obtained measuring the displacement of free aluminum beam since:

$$\Delta l_{al}^{free} = l_{al}^d - l_{al}^0 = l_{al}^d \alpha_{Si} \Delta T \quad (3)$$

Other methods are under investigation for measuring the displacement. The strain in the actuator beam is calculated using the same method using a term $\alpha\Delta T$ calibrated on free nitride beam. Then, assuming that the actuator strain is elastic, the strain and the stress in the actuator are related by Hooke's law (in uniaxial tension):

$$\sigma_{ac} = E_{ac}\varepsilon_{ac}^{el} \quad (4)$$

with $E_{ac}$ the Young modulus of the nitride. This modulus has been measured by nano-indentation and has been found to be equal to about 220 GPa which is in the range of the values usually found in the literature. The stress in the aluminum beam is then calculated by using the ratio between the surfaces:

$$\sigma_{al} = \sigma_{ac}\frac{S_{ac}}{S_{al}} \quad (5)$$

with $S_{ac}$ and $S_{al}$ the surfaces, respectively, of the actuator and of the aluminum. For small strains, the ratio $S_{ac}/S_{al}$ can be considered as constant. On each couple actuator beam/aluminum film a stress and a strain in the aluminum film are calculated. Thus a complete stress/strain curve is obtained using different initial lengths of actuator and aluminum beam as shown in Figure 4.

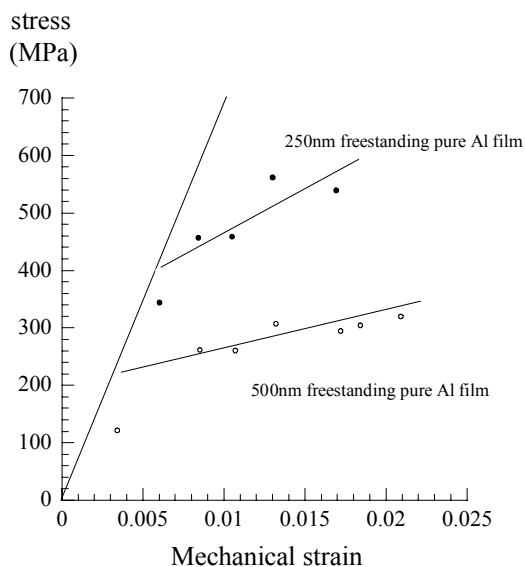

Figure 4: Stress strain curves obtained for pure Al films of various thicknesses.

In Figure 4, the elastic response of aluminum has been added by using a Young modulus equal to 70 GPa. Figure 4 shows the strong influence of the thickness of the Al film on the strength. Linear extrapolations have been carried out to evaluate the yield strength of the film. The yield strength is about 400 and 220 MPa for a 250 and 500 nm-thick film, respectively. The values of yield strength presented here are in the range of the values determined by other methods [15]. These values are much higher than the one usually found for pure bulk aluminum (usual value of yield strength for pure bulk aluminum at room temperature is about 70 MPa). This hardening of the material when considering thin films is due to complex dislocation mechanisms. The influence of the film thickness has already been underlined in the literature [15] and models for size-dependent plastic flow have already been studied [16].

### 3.2. Microstructural characterization of the thin films

The EBSD (Electron BackScattered Diffraction) technique has been used in order to determine the microstructure of deposited, aluminum film. The measurement of the grains orientation has been performed using a LEO Supra 55 SEM with a HKL Channel 5. The accelerating tension was set up to 20 kV and the sample was tilted to 70°. No sample preparation has been carried out. Figure 4 presents a map showing the different grains and their orientations perpendicular to the observation plane. The white parts in Figure 5 correspond to high disorientation points not indexed.

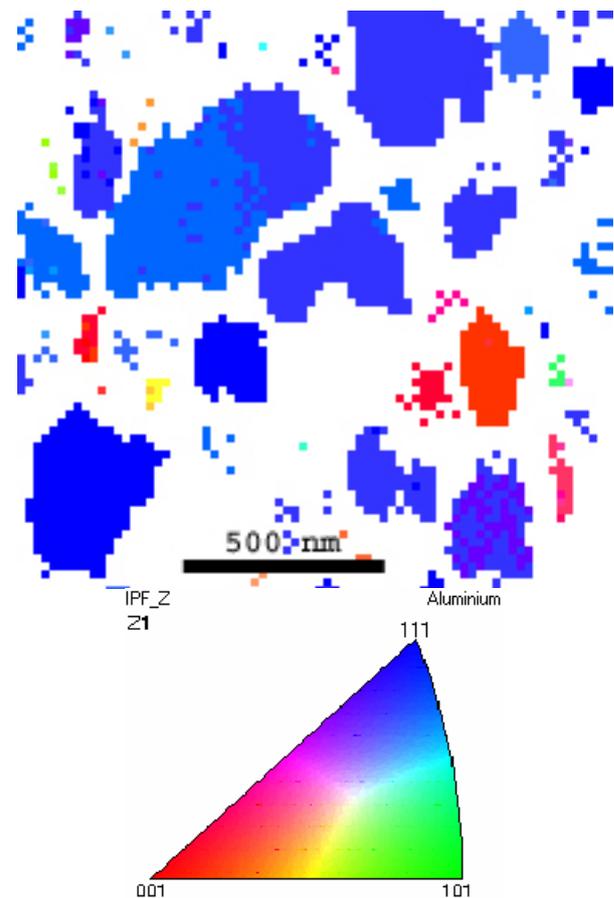

Figure 5: EBSD map showing the grains in a 250 nm aluminum film and their orientation.

Figure 5 shows that the aluminum grains have a rounded shape with a diameter in the range of 100 to 500 nm. The obtained grains are very small and this explain the high values of yield strength found in Section 3.1. when compared to literature results [15]. At last, all the grains are oriented in the same direction with the <111> axis perpendicular to the film surface. The EBSD method is a powerful tool to investigate the microstructure of thin deposited films.

## 4. CONCLUSIONS

The mechanical response of submicron aluminum films has been investigated using a new concept of on-chip micromachine. This micromachine uses the thermal stress resulting from the difference of thermal expansion coefficients between an actuator and the aluminum film to produce a displacement. Yield strength has been measured on films with two different thicknesses. The value of yield strength found as well as the increasing yield strength with the decreasing thickness of the film is in accordance with the literature results. EBSD measurements showed that the grain size is about 250 nm in a 250 nm-thick film and that all the grains are well oriented with their <111> axis in the thickness direction. This new concept of micromachine is promising and permits to obtain the mechanical properties of thin film in a straightforward way. It can be extended to other stress state, other materials and other films thicknesses. Some heat treatments can be carried out to change the grains size and to investigate the influence of the grain size on the mechanical properties of thin film for a given thickness. EBSD would permit to follow the microstructural changes involved by these kinds of heat treatments.

## 5. ACKNOLEDGEMENTS

The authors are grateful to Pr. P.J. Jacques from IMAP-UCL for the EBSD measurements. FRFC is also thanked. The support, through an ARC project, of UCL and Communauté Française de Belgique is gratefully acknowledged.